\documentclass[twocolumn,showpacs,preprintnumbers,amsmath,amssymb]{revtex4}
\usepackage{graphicx}
\usepackage{epsfig}
\usepackage{dcolumn}
\usepackage{bm}

\newcommand{\psip}{\psi(2S)}
\newcommand{\jpsi}{J/\psi}
\newcommand{\rar}{\rightarrow}
\newcommand{\rt}{\rightarrow}
\newcommand{\pipi}{\pi^+ \pi^- }
\newcommand{\etap}{\eta^{\prime}}
\newcommand{\piz}{\pi^0}

\newcommand{\kstark}{K^*(892)\overline{K}+c.c.}
\newcommand{\kst}{K^*(892)}

\newcommand{\pipipi}{\pi^+\pi^-\pi^0}
\newcommand{\gamg}{\gamma\gamma}

\def\Journal#1&#2&#3(#4){#1{\bf #2}, #3 (#4)}

\def\NPB{Nucl.  Phys.  {\bf B}}
\def\PLB{Phys.  Lett.  {\bf B}}
\def\PRL{Phys.  Rev.  Lett.  }
\def\PRD{Phys.  Rev.  {\bf D}}

\def\etal{{\it et al.}}

\def\bec{\begin{center}}
\def\eec{\end{center}}

\begin{document}
%\title{ $\psi(2S)\rightarrow Vector + Tensor$ }
\title{\boldmath Measurements of $\psi(2S)$ decays into
$\phi \pi^0$, $\phi \eta$, $\phi \eta^{'}$, $\omega
  \eta$, and $\omega \eta^{'}$ }
\author{M.~Ablikim$^{1}$, J.~Z.~Bai$^{1}$, Y.~Ban$^{10}$, 
J.~G.~Bian$^{1}$, X.~Cai$^{1}$, J.~F.~Chang$^{1}$, 
H.~F.~Chen$^{16}$, H.~S.~Chen$^{1}$, H.~X.~Chen$^{1}$, 
J.~C.~Chen$^{1}$
% ...
\\(BES Collaboration)\\ 
\vspace{0.2cm}
%\label{att}
$^1$ Institute of High Energy Physics, Beijing 100039, People's Republic of China\\
% ...
$^{10}$ Peking University, Beijing 100871, People's Republic of China\\
% ...
$^{16}$ University of Science and Technology of China, Hefei 230026, People's Republic of China\\
% ...
}

\begin{abstract}
  Decays of the $\psi(2S)$ into Vector plus Pseudoscalar meson final
  states have been studied with 14 million $\psi(2S)$ events collected
  with the BESII detector. Branching fractions of $\psi(2S)
  \rar\phi\eta$, $\phi\etap$, and $\omega\etap$, and upper limits of
  $\psi(2S) \rar \phi\piz$ and $\omega\eta$ are obtained: $B(\psi(2S)
  \rar\phi\eta) = (3.3\pm 1.1 \pm 0.5) \times 10^{-5}$, $B(\psi(2S)
  \rar\phi\etap) = (3.1\pm 1.4 \pm 0.7)\times 10^{-5}$, and $B(
  \psi(2S) \rar\omega\etap) = (3.2^{+2.4}_{-2.0} \pm 0.7) \times 10^{-5}$;
  and $B(\psi(2S) \rar\phi\piz) < 0.40 \times 10^{-5}$, and $B(\psi(2S)
  \rar\omega\eta) < 3.1 \times 10^{-5}$ at the 90 \% C.L..  These
  results are used to test the pQCD ``12\% rule''.
\end{abstract}
\maketitle

%\clearpage
\section{Introduction}

It is expected in perturbative QCD that both $\jpsi$ and $\psip$
decays to light hadrons proceed dominantly via three gluons or a
single virtual photon, with widths proportional to the squares of the
$c\overline{c}$ wave functions at the origin~\cite{qcd15}, which are
well determined from dilepton decays~\cite{PDG2004}. This led to the
``12\% rule'', i.e.
\begin{eqnarray*}
Q_h= \frac{B(\psip\rar h)}{B(\jpsi\rar h)}
   \simeq\frac{B(\psip\rar e^+e^-)}{B(\jpsi\rar e^+e^-)}\simeq 12\%.
\end{eqnarray*}
A strong violation of this conjecture was first observed by the
MarkII experiment in the Vector-Pseudoscalar meson (VP) final states,
$\rho\pi$ and $K^{*+}(892)K^{-}$~\cite{rhopi}. Significant suppressions
observed in four Vector-Tensor decay modes ($\omega f_2(1270)$, $\rho
a_2(1320)$, $\phi f_2^{\prime}$, and
$\kst\overline{K}^*(1430)+c.c.$)~\cite{BESII_VT} make the puzzle even
more mysterious.  Numerous theoretical explanations have been
suggested~\cite{qcd15_th}, but the puzzle still remains one of the most
intriguing questions in charmonium physics.
Recently both CLEO and BES reported new measurements of VP
channels~\cite{CLEOC_VP,BES_kstark, BES_rhopi} with higher statistics
and confirmed the severe suppression for $\rho\pi$ and $\kstark$.
%BES has considered interference effects, which have been
%neglected in almost all other analyses, in the
%analysis of the $\kstark$ and $\rho\pi$ channels.

In this letter, we report measurements of $\psi(2S)$ decays into 
5 VP channels: $\phi\piz$,
$\phi\eta$, $\phi\etap$, $\omega\eta$, and $\omega\etap$ using 14
million $\psip$ events collected with the BESII detector, where the 
branching fractions of  $\phi\etap$ and $\omega\etap$ are the first 
observations.  The 
results are compared with the corresponding $\jpsi$ branching fractions to
test the ``12\% rule''. Also, the branching fractions provide useful 
informations on DOZI suppressed decay $\psi(2S)\rar\phi\pi^0$ and on the 
quark components of $\eta$ and $\etap$~\cite{jdecay}.

\section{THE BESII DETECTOR}
The Beijing Spectrometer (BESII) is a conventional cylindrical
magnetic detector that is described in detail in Ref.~\cite{BES-II}.
A 12-layer Vertex Chamber (VC) surrounding the beryllium beam pipe
provides input to the event trigger, as well as coordinate
information.  A forty-layer main drift chamber (MDC) located just
outside the VC yields precise measurements of charged particle
trajectories with a solid angle coverage of $85\%$ of $4\pi$; it also
provides ionization energy loss ($dE/dx$) measurements which are used
for particle identification.  Momentum resolution of
$1.7\%\sqrt{1+p^2}$ ($p$ in GeV/$c$) and $dE/dx$ resolution for hadron
tracks of $\sim8\%$ are obtained.  An array of 48 scintillation
counters surrounding the MDC measures the time of flight (TOF) of
charged particles with a resolution of about 200 ps for hadrons.
Outside the TOF counters, a 12 radiation length, lead-gas barrel
shower counter (BSC), operating in limited streamer mode, measures the
energies of electrons and photons over $80\%$ of the total solid angle
with an energy resolution of $\sigma_E/E=0.22/\sqrt{E}$ ($E$ in GeV).
A solenoidal magnet outside the BSC provides a 0.4 T magnetic field in
the central tracking region of the detector. Three double-layer muon
counters instrument the magnet flux return and serve to identify muons
with momentum greater than 500 MeV/$c$. They cover $68\%$ of the total
solid angle.

In this analysis, a GEANT3 based Monte Carlo package with detailed
consideration of the detector performance (such as dead electronic
channels) is used.  The consistency between data and Monte Carlo has
been carefully checked in many high purity physics channels, and the
agreement is reasonable~\cite{J3pi}.
  
\section{Event Selection}

The data sample used for this analysis consists of $(14.0 \pm
0.6)\times10^6$ $\psip$ events~\cite{N_psip}, collected with BESII at
the center-of-mass energy $\sqrt s=m_{\psip}$.  The decay channels
investigated are $\psip$ into $\phi\piz$, $\phi\eta$, $\phi\etap$,
$\omega\eta$, and $\omega\etap$, where $\phi$ decays to $K^+K^-$,
$\omega$ to $\pi^+\pi^-\pi^0$, $\etap$ to $\eta\pi^+\pi^-$ or
$\gamma\pi^+\pi^-$, and $\pi^0$ and $\eta$ to 2$\gamma$. The events
have either two or four charged tracks plus $n$ ($n\geq 1$) photons.

\begin{figure}[h] \centering
\includegraphics[width=0.45\textwidth]{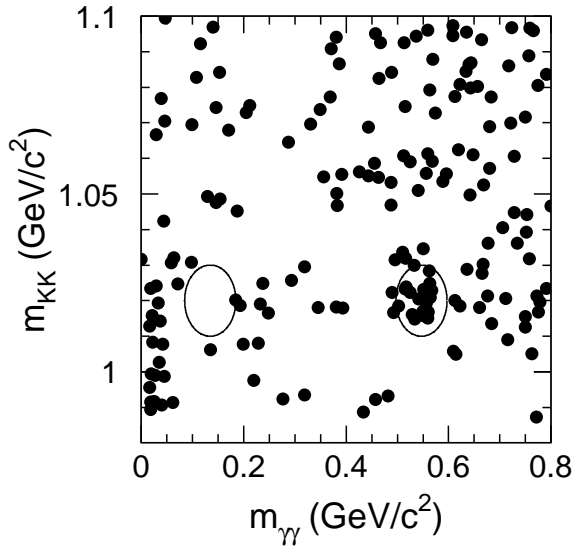}
\caption{\label{phigg}Scatter plot of $m_{KK}$ versus
   $m_{\gamma\gamma}$ for $\psip\rar K^+K^-\gamma\gamma$ candidate events.
The two ellipses in the plot indicate the $2\sigma$ contours of the signal regions.}
\end{figure}  

\begin{figure}[h] 
\includegraphics[height=5.5cm, width=0.4\textwidth]{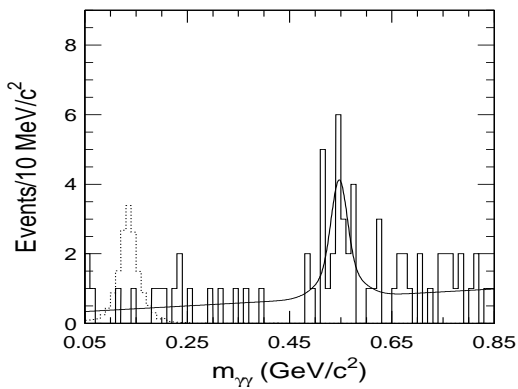}
\caption{\label{phigg_fit} Distribution of $\gamma\gamma$ invariant mass
  for $\psip\rar\phi\gamma\gamma$ candidate events (solid histogram)
  and Monte Carlo simulation of $\psip\rar\phi\pi^0$ (dotted
  histogram) with arbitrary normalization. The curve shows the best fit
  described in the text.}
\end{figure} 

A neutral cluster is considered to be a photon candidate if the
following requirements are satisfied: it is located within the BSC
fiducial region,
the energy deposited in the BSC is greater than 50 MeV, the
first hit appears in the first 6 radiation lengths, the angle in the
$x y$ plane (perpendicular to the beam direction) between the cluster
and the nearest charged track is greater than $16^\circ$ (this
requirement is not applied for channels involving more than two
photons), and the angle between the cluster development direction in
the BSC and the photon emission direction from the beam interaction
point (IP) is less than $37^\circ$.

Each charged track is required to be well fitted by a
three-dimensional helix, to have a momentum transverse to the beam
direction greater than 70 MeV/$c$, to originate from the IP region,
$V_{xy}=\sqrt{V_x^2+V_y^2}<2$ cm and $|V_z|<20$ cm, and to have a
polar angle $|\cos\theta|<0.8$. Here $V_x$, $V_y$, and $V_z$ are the
$x$, $y$, and $z$ coordinates of the point of closest approach of the
track to the beam axis.

The TOF and $dE/dx$ measurements for each charged track are used to
calculate $\chi^2_{PID}(i)$ values and the corresponding confidence levels
$Prob_{PID}(i)$
%($=Prob(\chi^{2}_{PID}(i),ndf_{PID})$)
for the hypotheses that
a track is a pion, kaon, or proton,
%where $ndf_{PID}=2$ is the number of degrees of freedom and
where $i$ ($i=\pi/K/p$) is the particle type.
For events with $\phi\rar K^+K^-$ decays, charged kaon candidates are
required to have $Prob_{PID}(K)$ larger than 0.01 or larger than
$Prob_{PID}(\pi)$ and $Prob_{PID}(p)$; while for events with
$\omega\rar\pipipi$ decays, at least half of the charged pion
candidates in each event are required to have $Prob_{PID}(\pi) >0.01$.

\subsection{\boldmath $\psip\rar\phi\piz$ and $\phi\eta$}
The $K^+K^-\gamma\gamma$ final state is utilized to measure these two
channels.  Two good charged tracks with net charge zero and at least
two photon candidates are required.  Next a four constraint (4C)
kinematic fit ($\chi^{2}_{kine}$) to the $K^+K^-\gamma\gamma$
hypothesis is performed, and the confidence level of the fit is
required to be larger than 0.01. If there are more than two photons,
the fit is repeated using all permutations of photons, and the two
photon combination with the minimum $\chi^{2}_{kine}$ is selected.
This procedure is used for all channels.  To suppress backgrounds from
$\pi/K/p$ misidentification, the combined chisquare~\cite{Chisquare},
$\chi^{2}_{comb}$, for the $\psip\rar K^+K^- \gamma\gamma$ assignment
is required to be smaller than those of $\psip\rar
\pi^+\pi^-\gamma\gamma$ and $\psip\rar p\overline{p}\gamma\gamma$.
Figure~\ref{phigg} shows the scatter plot of the invariant mass of
$K^+K^-$ ($m_{K^+K^-}$) versus that of the two gammas
($m_{\gamma\gamma}$).  A clear cluster can be observed in the
$\phi\eta$ signal region, while only one event appears in the
$\phi\piz$ region.

 To select $\phi$ decay candidates, we require $|m_{K^+K^-}-1.02|<0.02$
 GeV/$c^2$. Here the experiment mass resolution for $\phi$ is 2.5 MeV/$c^2$.
 The  ${\gamma\gamma}$ invariant mass distribution for events with
 $\phi$ candidates is shown in Fig.~\ref{phigg_fit}. 
 Fitting this distribution with $\pi^0$
 and $\eta$ functions determined from Monte
 Carlo simulation, plus an $\eta K^+K^-$ background determined by the 
 sideband and a first order polynomial to describe phase space
 background,
 $16.7\pm5.6$ $\phi\eta$ events are obtained with the statistical 
significance of $3.6\sigma$~\cite{significance}. While for the $\phi\pi^0$
channel, the observed events and the estimated background in the 
signal region are 4 and 6.2, respectively, which corresponds to the 
upper limit
 of 4.4 $\phi\pi^0$ events at the 90\% confidence level~\cite{upperlimit}.

\begin{figure}
\includegraphics[height=5.5cm, width=0.4\textwidth]{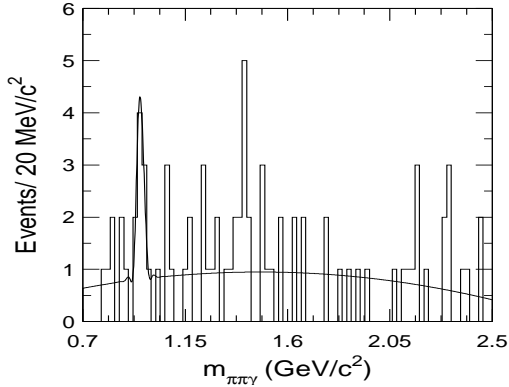}
\caption{\label{phietap2_fit}The $\gamma\pi^+\pi^-$ invariant mass
  distribution for $\psip\rar \phi\pi^+\pi^-\gamma$ candidate events.
  The curve shows the best fit described in the text.  }
\end{figure}

\begin{figure}
\includegraphics[height=5.5cm, width=0.4\textwidth]{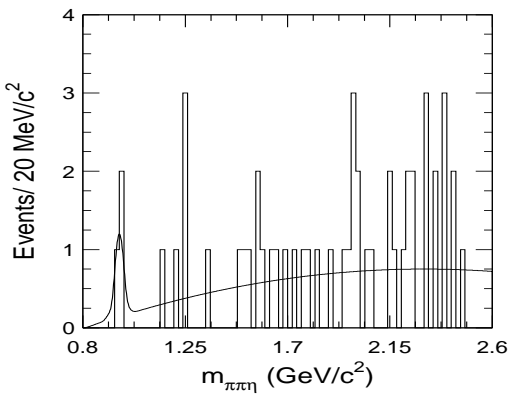}
\caption{\label{phietap_fit}The  $\pi^+\pi^-\eta$ invariant mass
  distribution for $\psip\rar \phi\pi^+\pi^-\eta$ candidate events.
  The curve shows the best fit described in the text.  }
\end{figure}

\subsection{\boldmath $\psip\rar\phi\etap$}
Two decay modes of $\eta^{\prime}(958)$ are used, $\eta^{\prime}(958) \rar
\pi^+ \pi^- \eta$ and $\gamma\pi^+\pi^-$.
Their final states are $K^+K^-\pi^+\pi^- +m\gamma$, where $m=1$ for
$\etap\rar\gamma\pi^+\pi^-$ and $m=2$ for $\etap \rar \pi^+ \pi^-
\eta$, $\eta \rar \gamma\gamma$.
Events with four charged tracks with net charge zero and at least $m$
photon candidates are selected. A 4C kinematic fit is performed for
the hypothesis $\psip\rar K^+K^-\pi^+\pi^- m\gamma$, and the
confidence level of the fit is required to be larger than 0.01. The
corresponding $\chi^2_{comb}$ for the $\psip\rar
K^+K^-\pi^+\pi^- m\gamma$ assignment is required to be smaller than those of
$\psip\rar \pi^+\pi^-\pi^+\pi^- m\gamma$ and $\psip\rar K^+ K^-K^+ K^-
m\gamma$.

The additional requirement $|m_{K^+
  K^-}-1.02|<0.02$ GeV/$c^2$, is used to select $\phi$ candidates, and
the $m_{\gamma\pi^+\pi^-}$ spectrum of selected events is shown in
  Fig.~\ref{phietap2_fit}. 
By fitting this spectrum with an
$\etap$ function, plus an $\etap K^+K^-$ background determined by the 
sideband and a second order polynomial for phase space 
background, $5.8\pm3.2$
$\phi\etap$ candidate events are obtained.
The statistical significance is about $2.0\sigma$.  Here, the shape of
$\etap$ is determined from Monte Carlo simulation of
$\psip\rar\phi\etap$, $\phi\rar K^+K^-$, and $\etap\rar
\gamma\pi^+\pi^-$.

For the final state $K^+K^-\pi^+\pi^-\gamma\gamma$, the
$K^+K^-\pi^+\pi^- \eta$ candidate events are required to satisfy
$|m_{\gamma\gamma}-0.547|<0.05$ GeV/$c^2$.  Backgrounds from
$\psip\rar \pi^{+}\pi^{-} \jpsi$, $\jpsi\rar \phi\eta$ and
$\psip\rar\eta\jpsi$, $\jpsi\rar\phi\pi^+\pi^-$ are eliminated with
two additional requirements $|m^{\pi^{+}\pi^{-}}_{recoil}-3.1|>0.1$
GeV/$c^2$ and $m_{K^+K^-\pi^+\pi^-}<2.9$ GeV/$c^2$, respectively.

With the requirement $|m_{K^+ K^-}-1.02|<0.02$ GeV/$c^2$, the
$\pi^+\pi^-\gamma\gamma$ invariant mass distribution for $\phi\pi^+\pi^-\eta$
candidate events is shown in Figure~\ref{phietap_fit}.  Fitting the
spectrum with an $\etap$ function plus a second order polynomial for
background, $2.6\pm1.8$ $\phi\etap$ candidate events are obtained.
The statistical significance is about $2.0\sigma$.  The $\etap$
shape is determined from Monte Carlo simulation of
$\psip\rar\phi\etap$, $\phi\rar K^+K^-$ and $\etap\rar
\pi^+\pi^-\eta$, $\eta\rar\gamma\gamma$.

\subsection{\boldmath $\psip\rar\omega\eta$}

\begin{figure}
\includegraphics[height=5.5cm, width=0.4\textwidth]{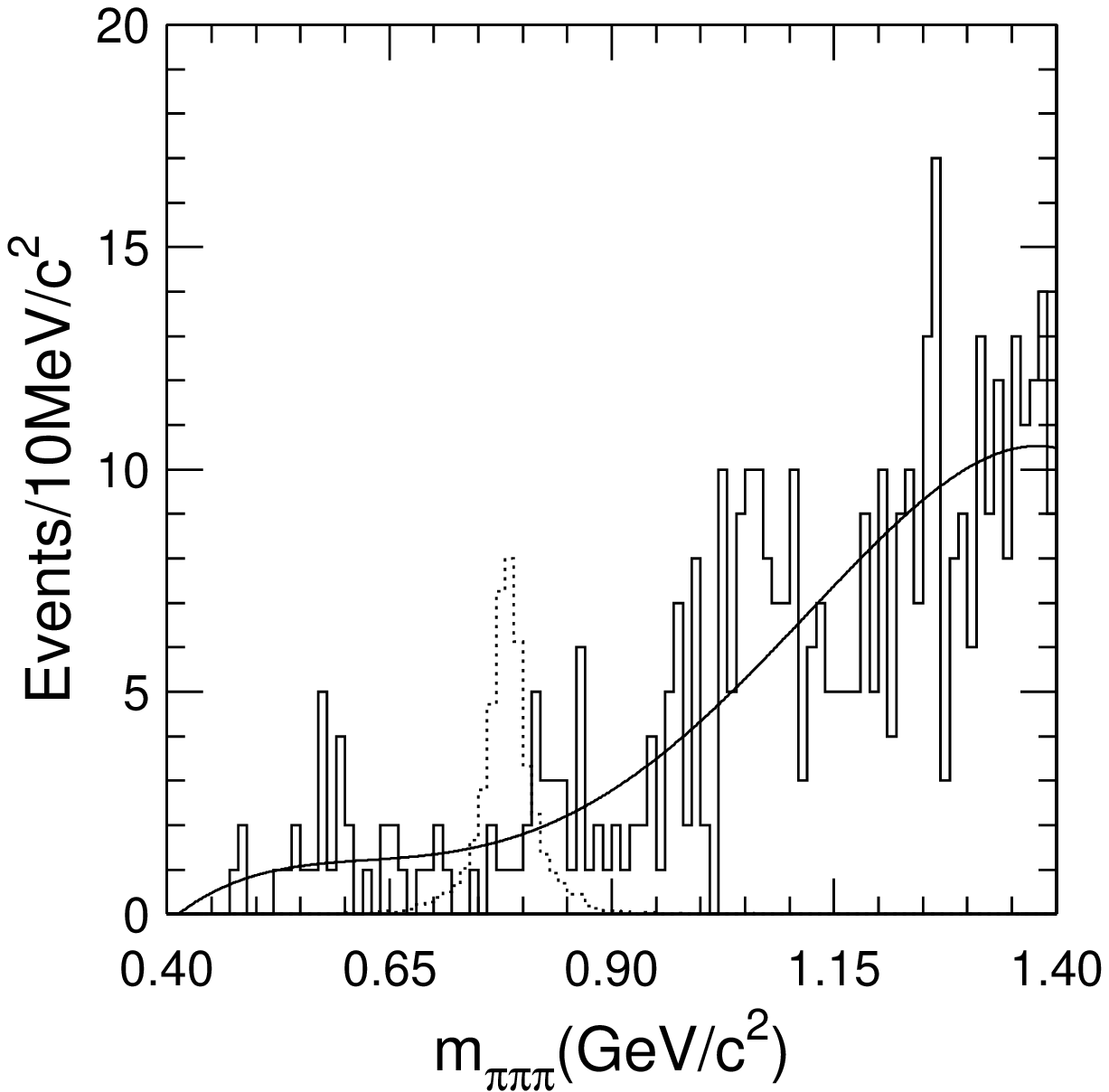}
\caption{\label{weta} The $\pipipi$ invariant mass distribution for
  $\pi^+\pi^-\pi^0\eta$ candidate events (solid histogram) and Monte
  Carlo simulation for $\psip\rar\omega\eta$ (dotted histogram) with
  arbitrary normalization. The
  curve shows the best fit described in the text.}
\end{figure}

Here, the final state studied is $\pipi\gamg\gamg$. Events with two
charged tracks with net charge zero and four or five photon candidates
are selected.  A 4C kinematic fit is performed for the hypothesis
$\psip\rar \pipi\gamg\gamg$, and the fit confidence level is required
to be larger than 0.01. To remove background from $\pi/K$
misidentification, $\chi^2_{comb}$ is required to be smaller
than that of $\psip\rar K^+K^-\gamma\gamma\gamma\gamma$.  Candidate
events must satisfy $|m_{\gamma_{1}\gamma_{2}} -0.135|<0.05$ GeV/$c^2$ and
$|m_{\gamma_{3}\gamma_{4}} -0.547|<0.05$ GeV/$c^2$ for the four photon
candidates ($\gamma_1\gamma_2\gamma_3\gamma_4$), where the subscripts
permute over all six combinations.  Events with one and only one
combination satisfying the above criteria are kept for further
analysis.

Figure~\ref{weta} shows the $\pipipi (m_{\pi\pi\pi})$ invariant mass
distribution after the above selection; no clear $\omega\eta$ signal
is seen.  The distribution is fitted with an $\omega$ signal determined 
by Monte Carlo simulation and a polynomial
background, the observed events and the estimated background in the 
signal region are 23 and 25.0, respectively, which corresponds to 
the upper limit of 9.7 $\omega\eta$ events at the 90\% confidence level.

\subsection{\boldmath $\psip\rar\omega\etap$}

\begin{figure}
\includegraphics[height=5.5cm, width=0.4\textwidth]{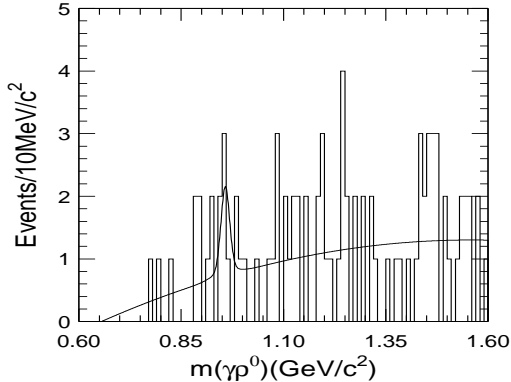}
\caption{\label{wetap-grho}The $\gamma\rho^0$ invariant mass distribution for
 $\psip\rar \pi^+\pi^-\pi^+\pi^-\pi^0\gamma$ candidate events. 
The curve shows the best fit described in the text. }
\end{figure}

\begin{figure}
\includegraphics[height=5.5cm, width=0.4\textwidth]{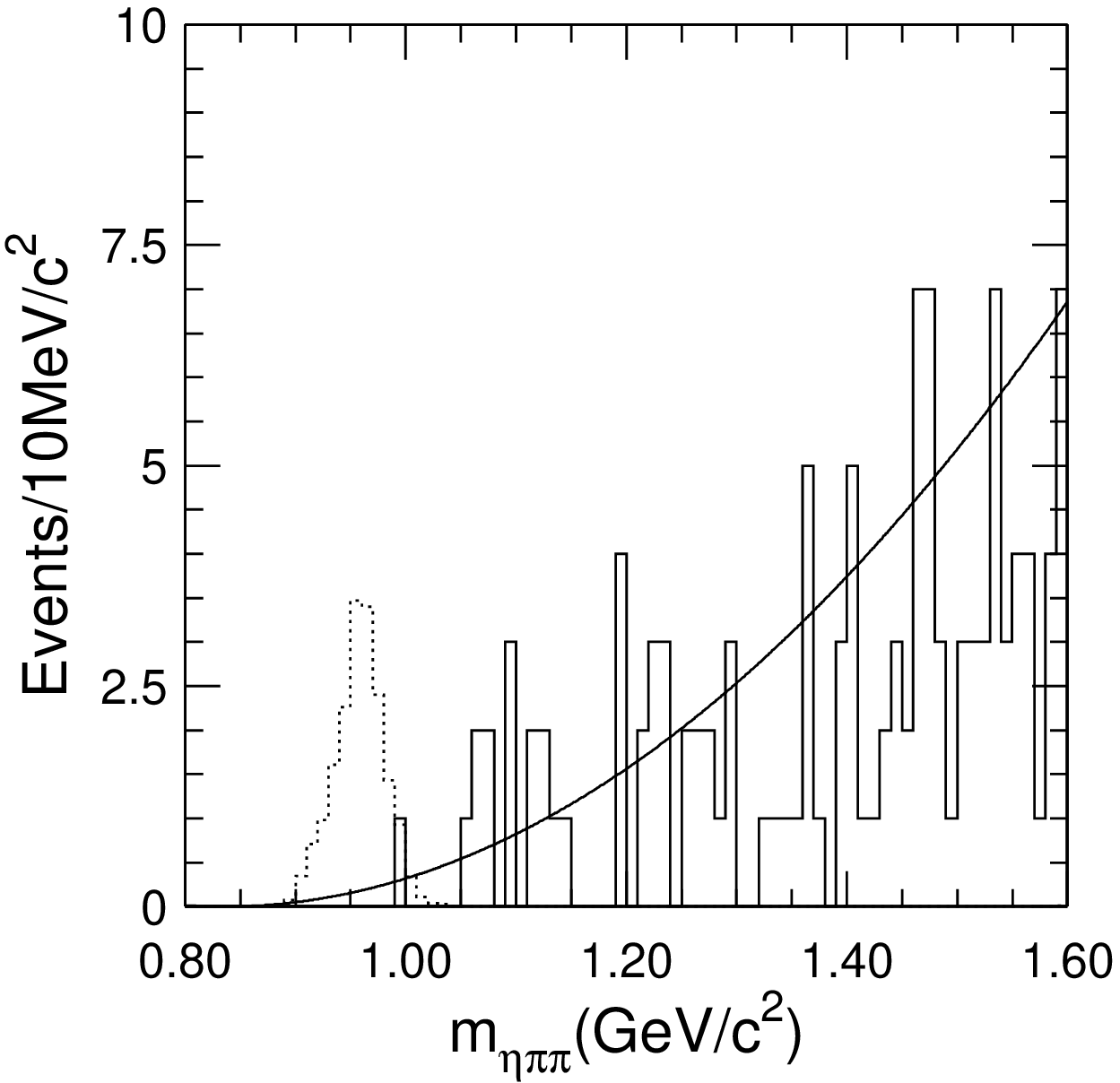}
\caption{\label{wetap}The $\eta\pipi$ invariant mass distribution for
  $\psip\rar \pipi\pipi\pi^0\eta$ candidate events (solid histogram)
  and Monte Carlo simulation of $\psip\rar\omega\eta^{'},
  \eta^{'}\rar \pi^+\pi^-\eta$ (dotted histogram) with arbitrary
  normalization.  The curve shows the best fit described in the text.
}
\end{figure}

Two $\eta^{\prime}(958)$ decay modes are used in this measurement,
similar to the measurement of $\psip\rar\phi\etap$. 
 Final states studied are $\pi^+\pi^-\pi^+\pi^- + m \gamma$, where
 $m=3$ for $\etap\rar\gamma\pi^+\pi^-$ and $m=4$ for
 $\etap\rar\eta\pi^+\pi^-$.
 Events with four charged tracks with net charge zero and $m$ or $m+1$
 photons candidates are selected. A 4C kinematic fit to the hypothesis
 $\psip\rar\pipi\pipi m\gamma$ is performed, and its confidence level
 is required to be larger than 0.01 and larger than that of $\psip\rar
 K^+ K^-\pipi m\gamma$ to suppress possible backgrounds due to
 particle misidentification.  Backgrounds from $\psip\rar\pipi J/\psi$
 are rejected with the requirement that the mass recoiling from every
 $\pipi$ pair
satisfy  $|m^{\pi^{+}\pi^{-}}_{recoil}-3.1|>0.05$ GeV/$c^2$.

For $\pi^+\pi^-\pi^+\pi^-\gamg\gamma$, one and only one pair among the
three good photon candidates is required to satisfy $|m_{\gamma\gamma}
-0.135|<0.05$ GeV/$c^2$; this pair is taken as a $\pi^0$.  To avoid
contamination from $\psip\rar\omega\pipi$, the $\pipipi\pipi$
invariant mass should be less than 3.5 GeV. Since the dominant decay
of $\eta^{'}$ into $\gamma \pi^+ \pi^-$ is $\gamma \rho$, an
additional requirement $|m_{\pi^+_1\pi^-_2} -0.771|<0.15 $ GeV is
applied to select $\pi^+\pi^-\pi^0\gamma\rho$ candidates, where
$\pi^+_1\pi^-_2$ is any combination from the four charged pion
candidates.

An additional requirement $|m_{\pipipi}-0.7826|<0.05$ GeV/$c^2$ is used to
select $\omega$ candidates.  The
$\gamma\rho^0$ invariant mass spectrum for selected events is shown in
Fig.~\ref{wetap-grho}.  It is fitted with an $\etap$
function determined by Monte Carlo simulation for
$\psip\rar\omega\etap$, $\etap\rar\gamma\rho$, plus a second order
polynomial for background, as shown in Fig. \ref{wetap-grho};
$4.2\pm2.7$ events are obtained with a statistical significance of
$1.9\sigma$.

For the final state $\pipi\pipi\gamg\gamg$, the selection of $\pi^0$
and $\eta$ is the same as for the $\omega\eta$ channel.  Only events
with only one $\pi^0$ candidate ($|m_{\gamg} -0.135|<0.05$ GeV/$c^2$)
and one $\eta$ candidate ($|m_{\gamg} -0.547|<0.05$ GeV/$c^2$) from
amongst the four photons are kept for further analysis.

An additional requirement $|m_{\pipipi}-0.7826|<0.05$ GeV is made to
select $\omega$ candidates. The $\eta\pipi$ invariant mass spectrum is
shown in Fig.~\ref{wetap}.  It is fitted with an $\etap$ function
determined by Monte Carlo simulation for $\psip\rar\omega\etap$,
$\etap\rar\eta\pipi$, plus a second order polynomial for background,
as shown in Figure \ref{wetap}. 
With  1 event observed in the signal region and 3.2 background events 
estimated from sidebands, the candidate $\omega\etap$ signal is 
$0_{-0}^{+1.7}$ event (at 68.3\% C.L.) assuming the Poisson 
variate~\cite{upperlimit}.
 
\section{Systematic errors}
Many sources of systematic error are considered. Systematic errors
associated with the efficiency are determined by comparing $J/\psi$
  and $\psi(2S)$ 
data and Monte Carlo simulation for very clean decay channels, such as
  $\psi(2S) \rt \pi^+ \pi^- J/\psi$, which
allows the determination of systematic errors associated with
the MDC tracking, kinematic fitting, particle
identification, and photon selection efficiencies~\cite{BESII_VT,J3pi}.

Since the decay $\eta^{\prime}$ to $\gamma\pipi$ includes
$\gamma\rho^0$ and  non-resonant $\gamma\pi^+\pi^-$, the $\pi^+\pi^-$ invariant
mass spectrum in the Monte Carlo simulation is obtained from
$J/\psi\rar \gamma\eta^{\prime}$, $\eta^{\prime}\rar\gamma\pi^+\pi^-$ data.
The uncertainty of their detection efficiency
from $\pi^+\pi^-$ invariant mass spectrum
is 3\%, which is included in systematic errors.

 The uncertainties of the branching fractions of intermediate
states, the background shapes in fitting, and the total number of $\psi(2S)$
events are also
sources of systematic errors.  Table~\ref{TSystematicError} summarizes
the systematic errors for all channels.

Contributions from the continuum $e^+ e^- \rightarrow
\gamma^*\rightarrow$ hadrons \cite{wangp} are estimated using a data
sample of $6.42\pm0.24$ pb$^{-1}$ taken at $\sqrt s=3.65$ GeV~\cite{3650},
corresponding to about one-third of the integrated luminosity at the 
$\psi(2S)$. Since no signal is observed for any channel analyzed,
the continuum contribution and possible interference 
are not taken into consideration.

\begin{table}[h]
\caption{\label{TSystematicError} Summary of relative
systematic errors ($\%$).}
%\mbox{} \hskip -2.5cm
\begin{tabular}{c|c|c|c|c|c}  \hline \hline
          &$\phi\pi^0$  & $\phi\eta$  & $\phi\etap$  
          &$\omega\eta$ & $\omega\etap$  \\ \hline
Tracking    &4.0 &4.0 &8.0  &4.0 &8.0  \\
$\gamma$ selection   &4.0 &4.0  &4.0 &8.0  &8.0 \\
Kinematic fit   & 6.0 & 6.0 & 6.0  &4.0 &4.0 \\
PID Efficiency      &2.0 &2.0 &2.0  & $\sim 0$  &$\sim0$\\
$\eta\rar\gamma\rho$ &-- &-- &1.9 &--  &1.9 \\
Background shape &0.0 &11.0 &16. &0.0  &18.9\\
MC statistics      &1.4 &2.1 &1.8 & 1.3& 2.0 \\
Branching ratios  &1.4 &1.6 & 3.8 &1.0  &3.5 \\ \hline
$N_{\psi(2S)}$      & \multicolumn{5}{c}{4.0} \\  \hline
Total                 &9.6 &15. &21. &11. &23. \\ \hline \hline
\end{tabular}
\end{table}

%%%%%%%%%%%%%%%%%%%%%%%%%%%%%%%%%%%%%%%%%%%%%%%%%%%%%%%%
\section{Results and Discussion}
The branching fraction for $\psi(2S)\rightarrow X$ is calculated from
$$B(\psi(2S)\rightarrow X)
  =\frac{n^{obs}_{\psi(2S)\rightarrow X\rightarrow Y}}
   {N_{\psi(2S)}\cdot B(X\rightarrow Y)\cdot \epsilon^{MC}},$$
where X is the intermediate state, Y the final state, and
$\epsilon^{MC}$ the detection efficiency.

Table~\ref{BESII_VP} summarizes the observed numbers of events,
detection efficiencies, and branching fractions or upper limits for
the channels studied.  
The branching fractions of $\psip\rar\phi\etap$ and
$\psip\rar\omega\etap$ are calculated from the sum of events observed
in the $\etap\rar\gamma\rho$ and $\eta\pi^+\pi^-$ channels
, and an efficiency determined from the
individual efficiencies weighted by the branching
fractions of these two channels.
  The upper limit for
$\psip\rar\omega\etap$ branching fraction at 90\% confidence level is 
$9.2\times 10^{-5}$. 
For comparison, the table includes
the corresponding branching fractions of $J/\psi$
decays~\cite{PDG2004}, as well as the ratios of the $\psi(2S)$ to
$J/\psi$ branching fractions.  Decays of $\psi(2S)$ to $\phi\eta$ and
$\omega\eta$ are suppressed by a factor of 2 and 12, respectively,
compared with the 12\% rule, while $\phi\etap$ and $\omega\etap$ are
consistent with the rule within large errors.  It is worth pointing
out that the ratio of $\frac{B(\phi\eta)}{B(\phi\etap)}$ is
$2.3\pm0.3$ and $1.1\pm 0.7$ for $\jpsi$ and $\psip$ decays,
respectively, which are consistent within 2$\sigma$, while the ratio of
$\frac{B(\omega\eta)}{B(\omega\etap)}$ is $9.5\pm1.7$ for
$\jpsi$ decay, which is much larger than that of $\psip$ decay.

This analysis without considering the continuum contribution
       (although not seen at present measurement) and possible
       interference might bring some uncertainty, which could only be
       clarified later by more accurate experiments such as CLEO-c or
       BESIII~\cite{BESIII}.
If the continuum contribution is treated incoherently, the continuum events,
assuming the Poisson distribution, for $\phi\eta$, $\phi\etap$ and 
$\omega\etap$ channels are $0^{+4.0}_{-0}$ at 68.3\% confidence level with
the normalized integrated luminosity, this yields the branching fractions
of $\psip \rar \phi\eta$, $\phi\etap$ and $\omega\etap$ to be
$(3.3^{+1.1}_{-1.4}\pm{0.5})\times 10^{-5}$,
 $(3.1^{+1.4}_{-2.0}\pm{0.7})\times 10^{-5}$ and
$(3.2^{+2.4}_{-3.2}\pm{0.7})\times 10^{-5}$, respectively.

{ %\small
\begin{table*}[hbt] 
\caption{\label{BESII_VP}
Branching fractions  and upper limits (90\% C.L.) measured for
$\psi(2S)\rightarrow$ Vector + Pseudoscalar. Results for corresponding
$J/\psi$ branching fractions \cite{PDG2004} and the
ratios $Q_h=\frac{B(\psi(2S)\rar h)}{B(J/\psi)\rar h)}$ are also given. } \vskip 2pt
\begin{tabular}{c|c|c|c|c|c}  \hline \hline 
$h$ & $N^{obs}$  & $\epsilon$ & $B(\psi(2S)\rar )$ & $B(J/\psi\rar)$ & $Q_h$ \\ 
         & &(\%) &  $\times 10^{-5}$ & $\times 10^{-4}$  & (\%) \\ \hline
$\phi\pi^0$      & $<4.4$ &16.1 & $<0.40$ & $<0.068$ & -- \\ 
$\phi\eta$       & $16.7\pm 5.6$ &18.9 & 
            $3.3\pm 1.1\pm0.5$ & $6.5\pm0.7$ & $5.1\pm 1.9$ \\
$\phi\etap(\etap\rar\gamma\pi^{+}\pi^{-})$ 
    & $5.8 \pm 3.2$ & 11.1 &  &  &  \\
$\phi\etap(\etap\rar\eta\pi^{+}\pi^{-})$  
    & $2.6 \pm 1.8$ & 3.8 &  &  &  \\
$\phi\etap(combined)$  & $8.4 \pm 3.7$ & 8.4 & 
      $3.1\pm 1.4\pm0.7$  & $3.3\pm0.4$ & $9.4 \pm 4.8$ \\
$\omega \eta$      & $<9.7$ & 6.3 & $<3.1$  & $15.8\pm1.6$ & $<2.0$  \\
$\omega\etap(\etap\rar\gamma\pi^{+}\pi^{-})$
    & $4.2 \pm 2.7$ & 2.6 &  &  &  \\
$\omega\etap(\etap\rar\eta\pi^{+}\pi^{-})$
    & $0.0_{-0.0}^{+1.7}$ & 1.8 &  &  &  \\
$\omega\etap(combined)$  & $4.2^{+3.2}_{-2.7}$ & 2.3 &
$3.2^{+2.4}_{-2.0}\pm0.7$ & $1.67\pm0.25$ & $19^{+15}_{-13}$ \\  \hline 
\hline 

\end{tabular} 
\end{table*}
}

 In conclusion, the branching fractions  for
$\psip\rar \phi\eta$, $\phi\etap$, and $\omega\etap$ and 
upper limits for $\phi\pi^0$ and $\omega\eta$ are presented.  
Our results for $\phi\etap$ and $\omega\etap$ are first measurements,
while our results for $\phi\pi^0$, $\phi\eta$, and $\omega\eta$
are consistent with those of CLEO~\cite{CLEOC_VP}.

\acknowledgments
   The BES collaboration thanks the staff of BEPC for their 
hard efforts. This work is supported in part by the National 
Natural Science Foundation of China under contracts 
Nos. 19991480, 10225524, 10225525, the Chinese Academy
of Sciences under contract No. KJ 95T-03, the 100 Talents 
Program of CAS under Contract Nos. U-11, U-24, U-25, and 
the Knowledge Innovation Project of CAS under Contract 
Nos. U-602, U-34 (IHEP); by the National Natural Science
Foundation of China under Contract No. 10175060 (USTC); 
and by the Department of Energy under Contract 
No. DE-FG03-94ER40833 (U Hawaii).


\begin{thebibliography}{99}
\bibitem{qcd15}T. Appelquist and H. D. Politzer, {\Journal\PRL&34&43(1975)};
               A. De Rujula and S. L. Glashow, {\em ibid}, page 46.
\bibitem{PDG2004} Particle Data Group, S. Eidelman \etal, {\Journal\PLB&592&1(2004)}, and references therein.
\bibitem{rhopi} M. E. B. Franklin \etal, MarkII Collab., {\Journal\PRL&51&963(1983).}
\bibitem{BESII_VT} J. Z. Bai \etal, BES Collab., {\Journal\PRL&81&5080(1998)}.
\bibitem{qcd15_th}W. S. Hou and A. Soni, {\Journal\PRL&50&569(1983)};
        G. Karl and W. Roberts, {\Journal\PLB&144&243(1984)};
        S. J. Brodsky \etal, {\Journal\PRL&59&621(1987)};
        M. Chaichian \etal, {\Journal\NPB&323&75(1989)};
        S. S. Pinsky, {\Journal\PLB&236&479(1990)};
        X. Q. Li \etal, {\Journal\PRD&55&1421(1997)};
        S. J. Brodsky and M. Karliner, {\Journal\PRL&78&4682(1997)};
        Yu-Qi Chen and Eric Braaten, {\Journal\PRL&80&5060(1998)};
        M. Suzuki, {\Journal\PRD&63&054021(2001)};
        J. L. Rosner, {\Journal\PRD&64&094002(2001)};
        J. P. Ma, {\Journal\PRD&65&097506(2002)};
        M. Suzuki, {\Journal\PRD&65&097507(2002)}.
%\bibitem{ichep97}Y. S. Zhu (Representing BES Collab.) in Proceedings of the 28th International  Conference on High Energy Physics, ed. Z. Adjuk and A. K. Wroblewski, World Scientific, 1997, p 507.
\bibitem{CLEOC_VP} N. E. Adam \etal, CLEO Collab., hep-ex/0407028.
\bibitem{BES_kstark}M. Ablikim \etal, BES Collob., hep-ex/0407037.
\bibitem{BES_rhopi} M. Ablikim \etal, BES Collob., hep-ex/0408047.
\bibitem{jdecay} L. K$\ddot{o}$pke and N. Wermes, Phys. Rep. {\bf 174}, 67 (1989)
\bibitem{BES-II} J. Z. Bai \etal, BES Collab., Nucl. Instr. Meth. {\bf A458},
 627 (2001).
\bibitem{J3pi} J. Z. Bai \etal, BES Collab., {\Journal\PRD&70&012005(2004)}.
%hep-ex/0402013, accepted by  Phys. Rev. {\bf D}.
\bibitem{N_psip} X. H. Mo \etal, HEP\&NP 28, 455(2004), hep-ex/0407055. 
\bibitem{Chisquare}
 The combined $\chi^2$, $\chi_{comb}^{2}$, is defined as the sum
of the $\chi^2$ values of the kinematic fit ($\chi^{2}_{kine}$) and those from
each of all particle identification assignments: 
$\chi_{comb}^{2}=\sum_{i}\chi^{2}_{PID}(i)+\chi^{2}_{kine}.$
\bibitem{significance}
The events number and statistical
significance of the $\phi\eta$ signal include the contribution
from $\psip$ decay and possibly from continuum, which is discussed   
in sections IV and V. This is also applicable for the $\phi\etap$ and
$\omega\etap$ channels.
The significance $S$ is calculated by $S=[-2\times 
(lnL_2-lnL_1)]^{1/2}$, 
where $L_1$ and $L_2$ is the likelihood function value in the fit with and 
without signal, respectively.
\bibitem{upperlimit} 
%M. Mandelkern and J. Schutz, J. Math. Phys. {\bf 41}, 5701(2000).
J. Conrad \etal,  {\Journal\PRD&67&012002(2003)}.
We use the modified likelihood ratio ordering including the systematic 
uncertainties of signal (Gaussian parametrization) and background (flat
parametrization) in confidence interval construction.
\bibitem{wangp} P. Wang, C. Z. Yuan, X. H. Mo and D. H. Zhang, {\Journal\PLB&593&89(2004)}; P. Wang, C. Z. Yuan and X.
H. Mo, {\Journal\PRD&69&057502(2004)}.
\bibitem{3650}S. P. Chi, X. H. Mo and Y. S. Zhu, HEP\& NP 28,1135(2004).
\bibitem{BESIII} Preliminary Design Report of The BESIII Detector, 
IHEP-BEPCII-SB-13, January, 2004.
\end{thebibliography}
\end{document}